\documentclass[10pt,a4paper]{article}
\pdfoutput=1

\usepackage[utf8]{inputenc}
\usepackage{uniinput}
\usepackage{jheppubmod}
\usepackage{amsmath}
\usepackage{url}
\usepackage{bm}
\usepackage{mathtools}
\usepackage{relsize}
\usepackage{cancel}
\usepackage{lipsum,cite,mathrsfs}
\usepackage[plain]{fancyref}
\usepackage{xcolor}

\definecolor{newred}{HTML}{DC3220}

\newcommand{\bea}{\begin{eqnarray}}
\newcommand{\eea}{\end{eqnarray}}
\newcommand{\be}{\begin{equation}}
\newcommand{\ee}{\end{equation}}

\usepackage{microtype}
\title{The dark bubbleography}
\author[a]{Souvik Banerjee,}
\emailAdd{souvik.banerjee@physik.uni-wuerzburg.de}
\affiliation[a]{Institut für Theoretische Physik und Astrophysik, Julius-Maximilians-Universität Würzburg,\\ Am Hubland, 97074 Würzburg, Germany}

\author[b]{Ulf Danielsson,}
\emailAdd{ulf.danielsson@physics.uu.se}
\affiliation[b]{Institutionen för fysik och astronomi,
	Uppsala Universitet, Box 803, SE-751 08 Uppsala, Sweden}

\author[a]{and Maximilian Zemsch}
\emailAdd{maximilian.zemsch@stud-mail.uni-wuerzburg.de}

\abstract{We present the holographic construction of the dark bubble model of dark energy and highlight the pivotal role played by the non-normalizable modes. Following the route of holographic renormalization, we show that the non-normalizable modes are essential for having a vanishing mass for the induced graviton in any braneworld model. We then apply this idea in the computation of the propagator on the wall of the dark bubble introduced in \cite{Banerjee:2018qey}. }

\preprint{UUITP-33/23}

\begin{document}

\maketitle

\section{Introduction and overview}\label{sec:intro}
The dark bubble model, first proposed in \cite{Banerjee:2018qey} and subsequently developed in \cite{Banerjee:2019fzz, Banerjee:2020wix, Banerjee:2020wov, Banerjee:2021qei, Banerjee:2021yrb, Banerjee:2022myh,Danielsson:2021tyb, Danielsson:2022lsl,Danielsson:2022fhd,Basile:2023tvh,Danielsson:2023alz}, is an alternative to standard string compactifications, which potentially circumvents the swampland obstructions against obtaining a stringy model of dark energy. In this paper, we will focus on presenting a complete and consistent holographic description of this model in terms of a necessary generalization of techniques used in the AdS/CFT correspondence \cite{Maldacena:1998im, Witten:1998qj, Gubser:1998bc}. In particular, we will sort out how the choice of holographic boundary conditions in our model is related to the mass of the 4D graviton imprinted on the bubble wall.

Similarly to the Randall-Sundrum scenario \cite{Randall:1999vf}, the dark bubble model makes use of branes embedded into a five-dimensional AdS space, even though there are some crucial differences. The Randall-Sundrum model embeds the universe as a hypersurface between two higher-dimensional AdS spaces. The extrinsic curvatures on the two sides of the embedded brane have the same sign, such that the two spaces both correspond to insides with respect to the braneworld. It is also common to impose a $\mathbb{Z}_2$ symmetry, identifying the two bulk spacetimes. The dark bubble is geometrically very different since the hypersurface is given by the spherical shell of a bubble of true vacuum expanding inside a decaying metastable AdS vacuum in the exterior. Clearly, there is no $\mathbb{Z}_2$ symmetry, but even more important is the fact that the bubble has an inside and an outside.

The imbalance between the inner $(-)$ and outer $(+)$ space of the bubble is reflected in different values of the bulk cosmological constants $\Lambda_\pm=-6k_\pm^2$, with $k_- > k_+$, on the two sides of the bubble wall. The cosmological constant outside of the shell is larger than the one inside of the shell, which induces the decay of the false vacuum to the true one via the nucleation of a brane. The metric is of the form
\begin{align}
    \mathrm{d}s^2=-f_\pm(r) \, \mathrm{d}t^2+\frac{1}{f_\pm(r)}\, \mathrm{d}r^2 + r^2 \, \mathrm{d} \Omega_{3}^2
\label{AdSmetric}
\end{align}
with $f_\pm(r)=1+k_\pm^2r^2$ for the interior $(-)$ or exterior $(+)$, respectively. The dynamical expansion parameter $a(t)$ determines the location of the shell, so that $r<a(t)$ holds for the inside and $r>a(t)$ for the outside of the bubble. For an observer on the bubble wall, a proper time $\tau$ can be chosen in such a way that the shell metric recovers the geometry of a FLRW universe. The proper time evolution of the bubble radius $r=a(\tau)$ is determined by the Israel junction conditions, which require
\begin{align}
\sigma = \frac{3}{8\pi G_5} \bigg( \sqrt{k_-^2+\frac{1+\dot{a}^2}{a^2}}-\sqrt{k_+^2+\frac{1+\dot{a}^2}{a^2}}\bigg) ,
\label{tension}
\end{align}
where $\sigma$ is the tension of the shell.
Here, $\dot{a}=\mathrm{d}a/\mathrm{d}\tau$. The critical value of the brane tension is given by
\begin{align}
\sigma_\text{crit} := \frac{3}{8\pi G_5}(k_- - k_+),
\end{align}
and corresponds to a flat Minkowski universe with vanishing cosmological constant. In case of the dark bubble, the tension stays slightly subcritical, meaning $\sigma=\sigma_\text{crit}(1-\epsilon)$ with a small and positive parameter $\epsilon$. With that, equation (\ref{tension}) can be expanded in powers of $\epsilon$ yielding the Friedmann equation in four dimensions
\begin{align}
H^2\equiv \frac{\dot{a}^2}{a^2} = - \frac{1}{a^2} + \frac{8\pi G_4}{3}\Lambda_4 + \mathcal{O}(\epsilon^2), 
\label{Friedmanneq}
\end{align}
where
\begin{equation}
    \Lambda_4 = \sigma_\text{crit} -\sigma >0
\label{ccbubble}
\end{equation}
is the 4D cosmological constant, associated with a positive energy density. As described in \cite{Banerjee:2018qey}, and reviewed in \cite{Banerjee:2019fzz}, the 4D Newton's constant is given by 
\begin{equation}
\label{eqn:G4-DB}
    G_4 = \frac{2 k_- k_+}{k_- - k_+} G_5 .
\end{equation}
One can also consider more general asymptotically AdS spacetimes, e.g. AdS-Schwarzschild geometry, which induces an additional term to the Friedmann equation \eqref{Friedmanneq} that can be identified as radiation on the bubble wall \cite{Banerjee:2018qey}. Similarly, a spherically symmetric cloud of strings radially stretched in the bulk can potentially induce four-dimensional matter on the wall \cite{Banerjee:2019fzz}.  

Contrary to RS, neither gravity nor matter is localized on the brane. The uplift of 4D gravitational waves to 5D has been explored in detail in \cite{Danielsson:2022fhd}. There, it was demonstrated how the dark bubble sustains 5D gravitational waves, which induces the expected 4D waves in metric on the brane. Furthermore, in \cite{Basile:2023tvh}, it was shown how electromagnetic waves, identified with the excitations of the gauge field within the brane, source the string theory Kalb-Ramond $B$-field in the bulk. The backreaction of these fields on the 5D bulk metric induces the expected 4D gravitational backreaction. What these works explicitly show, which is further reviewed in, e.g. \cite{Banerjee:2022myh}, is that the matter fields attached to the brane are governed by dynamics that is constrained to that of 4D Einstein gravity coupled to matter. This follows from the Gauss-Codazzi equations, which express the induced metric in terms of the extrinsic curvature and the bulk metric, together with the 5D Einstein equations and the junction conditions. As explained in \cite{Basile:2023tvh}, the spreading out of matter and gravity in the fifth dimension only show up as deviations in the Einstein equations at high energy densities.\footnote{Just as in the case of RS, there are also modifications in the force of gravity at distances of the order of the AdS-scale. Note that this scale is always assumed to be microscopic. For further discussions on this see \cite{Basile:2023tvh,Danielsson:2023alz}.}  

The case of RS is superficially similar, but fundamentally different. The key difference is the junction condition \eqref{tension}, where the sign of the second term of the junction condition is changed from a minus to a plus since there are two insides. Considering the $\mathbb{Z}_2$-symmetric case for simplicity, this leads to an effective 4D cosmological constant given by $\sigma - 2k$, and the 4D Newton's constant given by $G_4 =k G_5$. The changes in the phenomenology these differences bring about are profound, as elaborated in \cite{Banerjee:2020wov, Banerjee:2022myh}. In particular, there is now a {\it lower} bound on $\sigma$ to obtain a positive cosmological constant. Contrary to the case of the dark bubble, extremal branes are far from the bound, which naturally leads to an AdS braneworld rather than a universe with positive dark energy. We will not discuss RS any further in the present paper, but instead focus on how the phenomenologically promising dark bubble model can be embedded, consistently, in a holographic setting and the consequences thereof.

It turns out that this key difference between the Randall-Sundrum braneworld model and our dark bubble becomes pivotal to ensure a consistent holographic description of the latter. Since the bubble has an inside and an outside, it can naturally define a monotonically increasing holographic direction, contrary to the case of Randall-Sundrum, where a holographic description requires a mirroring of the holographic direction across the brane. Therefore, the applicability of holography is even more manifest in our dark bubble model. This is what we aim to elaborate on in this paper.

Massive particles can be represented by strings that end on the bubble wall and extend in the radial direction. They represent matter fields in the expanding universe whose masses are effectively renormalized by virtue of the holographic renormalization of the Newton constant $G_4$ \cite{Banerjee:2020wov}. The energy carried by the endpoints of the strings is associated with non-normalizable metric modes on the boundary hypersurface. However, there is no obvious reason why non-normalizable modes should be necessary or even wanted in a holographic description. Here, we show that this follows quite naturally from the requirement of a massless graviton in a quantum description of gravity, induced on the bubble wall. {\it The present work presents a systematic and consistent holographic approach to elucidate the pivotal role of bulk non-normalizable modes in obtaining a massless graviton on the bubble wall. These non-normalizable modes, in the context of the dark bubble model, are manifestations of string sources which eventually provide the energy for the expansion of the bubble \cite{Banerjee:2020wov}.} A very similar role of the non-normalizable modes in the context of the Karch-Randall braneworld model \cite{Karch:2000ct} has been discussed in a recent work \cite{Neuenfeld:2021wbl}. Our conclusion is very much in line with that.

In the holographic setup, variation of the gravity action in AdS$_5$ yields a boundary contribution of the form
\begin{equation}
\label{var1}
\delta S_5 = \frac{1}{2} \int_{\partial\mathcal{M}} d^4x \sqrt{-\gamma} T_{\mu\nu}^{\mathrm{(CFT)}} \delta \gamma^{\mu\nu},
\end{equation}
where $\gamma^{\mu\nu}$ is the induced metric on the cutoff hypersurface $r=a$ and $\gamma$ is the determinant of $\gamma^{\mu\nu}$ on that hypersurface. $T_{\mu\nu}^{\mathrm{(CFT)}}$ is the bare stress-energy tensor of the dual conformal field theory living on 
the $r=a$ hypersurface. Conventionally, in this holographic setup one tends to employ Dirichlet boundary 
conditions on the boundary hypersurface, which fix the boundary metric. As a consequence, \eqref{var1} vanishes leading to a well-defined variational principle. Nevertheless, by construction, it does not allow for any dynamical boundary metric. On the contrary, having a dynamical metric on the bubble wall is a necessity in our model in order to reproduce an expanding universe.
From \eqref{var1} it is therefore clear that in order to achieve a dynamic metric on the wall, we do need to go for different boundary conditions that can be imposed without requiring to fix the metric on the 
boundary. One possibility to obtain such a dynamic geometry on the boundary has been investigated in the context of holographic cosmology \cite{Banerjee:2012dw, Apostolopoulos:2008ru}, where it was shown that a ``mixed" boundary condition instead of a usual Dirichlet can potentially yield  cosmological evolution on the boundary. This mixed boundary condition, which amounts to adding a suitable local action on the boundary, falls into the allowed class of boundary conditions in AdS/CFT \cite{Compere:2008us}. We show in this paper that this particular choice of boundary condition is crucial to obtain a holographic realization of the dark bubble model.

The structure of this paper is as follows: To set the stage, in section \ref{sec:holocos} we start with a lightning review of the holographic approach to cosmology, emphasizing the importance of choosing an appropriate boundary condition to obtain expanding cosmology on a cutoff near the boundary of an asymptotically AdS spacetime. This involves a holographic renormalization of the induced Newton's constant, and the cosmological constant on the cutoff brane. In section \ref{sec:NN}, we illustrate the role of normalizable modes in inducing a massive graviton in the simple setup of holographic cosmology discussed in section \ref{sec:holocos}. In this model, there is only one AdS bulk, which we generalize, in  section \ref{sec:bubble}, to accommodate our dark bubble. Apart from having an inside and an outside, this model also requires consideration of the dynamics of the brane source. Incorporating both these features, and adopting the lessons learnt from previous sections, we present the computation of the graviton propagator on the bubble wall. Our computation elucidates the precise role of non-normalizable modes towards obtaining the dynamics of a massless graviton on the wall of the dark bubble. We conclude our short report in section \ref{sec:conclusion}.

\section{Holographic Cosmology - a primer}
\label{sec:holocos}
In this section we present
the central ingredients that are essential for building up an expanding cosmological model on a holographic screen. The holographic screen is here a UV cutoff near the boundary of an asymptotically AdS spacetime which can as well be a codimension-1 brane with a fixed value of tension. There had been earlier attempts to realize cosmology in the realm of AdS/CFT, including a surge of research activities to establish a holographic description of RS braneworld models \cite{Gubser:1999vj, Kraus:1999it, Barcelo:2000re, Anchordoqui:2000du,Savonije:2001nd, Hebecker:2001nv, Padilla:2001jz,Shiromizu:2001jm, Gregory:2001dn, Gregory:2002am,Biswas:2001sh, Cai:2001jc, Petkou:2001nk, Mukherji:2002ft, Maartens:2003tw, Kiritsis:2005bm}.
However, as discussed above, there was still an outstanding puzzle regarding the choice of appropriate boundary conditions that can potentially give rise to a dynamical cosmological spacetime on the boundary. As discussed above, this was due to the fact that the Dirichlet boundary condition used in standard AdS/CFT renders the boundary metric fixed and time-independent, which was clearly not conducive for realizing expanding cosmology at the boundary. This issue was successfully addressed in \cite{Apostolopoulos:2008ru, Banerjee:2012dw} which we briefly gloss over below. 

Our setup consists of a $d$-dimensional de Sitter hypersurface living close to the boundary of an $\text{AdS}_{d+1}$ spacetime, which we will later promote to a holographic model of a dark bubble.
The metric of the bulk is of the form 
\begin{align}
\label{eqn:metric-global}
    \mathrm{d}s^2=-f(r) \, \mathrm{d}t^2+\frac{1}{f(r)}\, \mathrm{d}r^2 + r^2 \, \mathrm{d} \Omega_{d-1}^2 ,
\end{align}
with the radial function $f(r)$. The spacetime is cut off by a brane at a radial position $a$ that evolves in time. This is a crucial requirement since the boundary hypersurface in this setup corresponds to the expanding universe and hence has to be dynamical. It is convenient to introduce a proper time parameter $\tau$ as the brane time, and then to choose a suitable parametrization $\{t(\tau),a(\tau)\}$ in order to adjust the boundary metric appropriately. For cosmological applications, we want to imprint an FRW metric on the boundary:
\begin{align}
 \mathrm{d}s^2=-\mathrm{d}\tau^2 + a^2(\tau) \mathrm{d}\Omega_{d-1}^2.
\end{align}
In order to describe the dynamics of the hypersurface, we now need to impose suitable boundary conditions. As argued above, in order to ensure real-time dynamics on the cutoff surface, which can as well be thought of as an end-of-the-world  brane with fixed tension, we need to dump the Dirichlet conditions, and instead  resort to mixed boundary conditions. In order to implement this boundary condition, we need an appropriate local stress-energy tensor $T_{ij}^\mathrm{local}$ along with the induced CFT stress-energy tensor $T_{ij}^\mathrm{CFT}$  on the brane. $T_{ij}^\mathrm{CFT}$ is obtained by varying the full bulk action supplemented by the Gibbons-Hawking-York counter term \cite{Balasubramanian:1999re}. Furthermore, to render the full stress-energy tensor finite, even in the limit when the expansion parameter goes to infinity $a(\tau) \rightarrow \infty$, we need to add, additionally, a counterterm $T_{ij}^\mathrm{ct}$ which is obtained by varying the counterterm action
\begin{align}
S_\mathrm{ct} = - \frac{1}{2} \int \limits_{\partial \mathcal{M}} \mathrm{d}^d x \sqrt{-\gamma} (\kappa_1 \mathcal{R}[\gamma] + \kappa_2),
\label{countertermaction}
\end{align}
that appears in the bulk action following holographic renormalization\cite{deHaro:2000vlm}.
Compiling all these ingredients, the mixed boundary condition reads \cite{Compere:2008us}
\begin{align}
T_{ij}^\mathrm{CFT} + T_{ij}^\mathrm{local} + T_{ij}^\mathrm{ct} = 0.
\label{mixedBC}
\end{align}
The different components appearing in  \eqref{mixedBC} take the following explicit expressions,
\begin{align}
 T_{ij}^\mathrm{CFT} &= \frac{1}{8\pi G_{d+1}}(K_{ij} - K \gamma_{ij}), \label{TCFTdef}\\ 
 T_{ij}^\mathrm{local} &= -\frac{1}{8\pi G_{d}}\bigg({\cal R}_{ij}-\frac{1}{2}{\cal R}\gamma_{ij}+\Lambda_{d} \gamma_{ij}\bigg),\\
 T_{ij}^\mathrm{ct} &= -\kappa_1 \bigg({\cal R}_{ij}-\frac{1}{2}{\cal R}\gamma_{ij}\bigg) -\kappa_2 \gamma_{ij}.
\end{align}
Here, the Ricci tensor ${\cal R}_{ij}$ and the Ricci scalar $\cal R$ are defined on the brane, so is the extrinsic curvature $K_{ij}=1/2 \,
n^k\partial_k \gamma_{ij}$. The parameters $\kappa_1, \kappa_2$ are determined through holographic renormalization \cite{deHaro:2000vlm}. Technically, these parameters are chosen such that the energy density $\epsilon$ and the pressure $p$ on the cutoff brane are finite. In a 4+1 dimensional asymptotically AdS spacetime, the energy density, pressure and the trace of the stress-energy tensor on the brane are given by \cite{Banerjee:2012dw}
\begin{align}
\label{eqn:epsilon-p}
 \epsilon &= T_{\tau \tau}^\mathrm{CFT} + T_{\tau \tau}^\mathrm{ct} = \kappa_2 + \kappa_1 \bigg( H^2 + \frac{\tilde k}{a^2} \bigg) - \frac{3}{8\pi G_5 a} \sqrt{\dot{a}^2 + f(r)}, \nonumber \\
 p &= T_{i}^{i, \, \mathrm{CFT}} + T_{i}^{i, \, \mathrm{ct}} = - \kappa_2 - \kappa_1 \bigg(H^2 + \frac{\tilde k}{a^2} + \frac{2\ddot{a}}{a} \bigg) + \frac{1}{16\pi G_5} \frac{a f' +2(a\ddot{a}+2\dot{a}^2) + 4f}{a\sqrt{\dot{a}^2+f}},\nonumber \\
{\rm Tr}\,T &= - \frac{3L^3}{16 \pi G_5} \bigg(H^2 + \frac{1}{a^2} \bigg) \frac{\ddot{a}}{a} + \mathcal{O} (H^4),
\end{align}
with $H=\dot{a}/a$ being the Hubble parameter and $\tilde k = 0,\pm 1$, depending on whether the asymptotic boundary is flat, spherical or hyperbolic, respectively.
In order to render \eqref{eqn:epsilon-p} finite in the limit $a\rightarrow \infty$, one needs to choose
\begin{align}
\label{eqn:kappa12}
\kappa_1=\frac{3L}{16\pi G_5} \, \, \, \,   \text{and} \, \, \, \, \kappa_2=\frac{3}{8\pi G_{5}L},  
\end{align}
which can be thought of as renormalizing the bare 4-dimensional Newton's constant and the 4-dimensional cosmological constant, respectively.

Having a non-trivial dynamic metric on the boundary will nevertheless break the conformal invariance on the boundary, which can be tracked, systematically, by computing the Weyl anomaly. To demonstrate this, let us consider a 4+1 dimensional AdS Schwarzschild black hole with the blackening factor $f(r) = {\tilde k} +\frac{r^2}{L^2}- \frac{M}{r^2}$ with a flat static boundary so that $H = {\tilde k} = 0$. In this case, the renormalized energy density and pressure on the brane simplifies as an expansion in large $a$ as
\begin{align}
\epsilon = \frac{3}{8\pi G_5 L} \bigg(\frac{(\pi LT)^4}{2}-\frac{7(\pi L T)^8}{8} + ...\bigg), \, \, \,
p=\frac{1}{8\pi G_5 L} \bigg( \frac{(\pi L T)^4}{2} - \frac{3(\pi L T)^8}{8} + ... \bigg),
\end{align}
with $T=T_\mathrm{Hawking}/\sqrt{\dot{a}^2+f(a)}$ being the red-shifted Hawking temperature on the cutoff brane. Clearly, $\epsilon=3p \propto T^4$ holds at the leading order of $a$-expansion which is thermodynamically expected for a conformal fluid. This conformality is broken when the subleading contributions are considered, i.e. when we move a finite distance away from the boundary and is broken further when the full dynamics of the brane with $H\ne0$ is considered. The latter is evident from the non-vanishing trace term $\text{Tr} \, T = \epsilon - 3p$ as in \eqref{eqn:epsilon-p}.
Nevertheless, \eqref{eqn:epsilon-p} still satisfies the first law of  thermodynamics at the leading order of $a$-expansion,
\begin{align}
\mathrm{d}E = T \mathrm{d}S - p \mathrm{d}V,
\end{align}
$S=(\pi L T)^3 V^{(3)}/(4G_5)$ being the entropy of the black hole, $E=\epsilon V$ and $V=a^3 V^{(3)}$ where $V^{(3)}$ is the three-dimensional transverse volume. 
Finally, plugging \eqref{eqn:epsilon-p} and \eqref{eqn:kappa12} back in the $\tau\tau$ component of \eqref{mixedBC} yields the cosmological evolution in the form of a Friedmann equation on the brane, in the leading order of expansion in $1/a$. 
\begin{align}
H^2\equiv \frac{\dot{a}^2}{a^2} = - \frac{\tilde k}{a^2} + \frac{8\pi G_4}{3}\epsilon + \frac{1}{3} \Lambda_4. 
\label{Friedmanneq1}
\end{align}

\section{Non-normalizable modes and the graviton mass in a braneworld construction} \label{sec:NN}
In the simple holographic setup discussed above, in this section, we will derive a formula for graviton mass induced on the cutoff brane.
For this purpose, we start with a Fefferman-Graham expansion of an asymptotically  $\text{AdS}_{d+1}$ bulk metric \eqref{eqn:metric-global} which assumes a generic form \cite{Skenderis:2002wp}

\begin{align}
\mathrm{d}s^2= \frac{L^2}{4\rho^2} \mathrm{d}\rho^2 + \frac{L^2}{\rho} g_{ij}(x, \rho) \mathrm{d}x^i \mathrm{d}x^j, 
\end{align}
where $g_{ij}(x,\rho)$ assumes a power series expansion near the boundary, 
\begin{align}
\label{eqn:FG-exp}
    g(x,\rho) = g_{(0)} (x)+ \cdots g_{(d)}(x) \rho^{\frac{d}{2}} + {\bar g}_d (x) \,\rho^{\frac{d}{2}} \log \rho,
\end{align}
$g_{(0)}$ being the boundary metric. The last term is an anomaly term which contributes only for even boundary dimensions. $L$ is the AdS radius and $\rho$ is the new radial coordinate with $\rho \rightarrow 0$ being the asymptotic boundary. As described in the previous section, we introduce a brane at the hypersurface close to the boundary of AdS at $\rho = \epsilon$ where $\epsilon$ serves as a UV cutoff from the perspective of the dual field theory. In order to have a fully consistent boundary theory without any UV divergence, one would naively expect the modes of the induced metric $\gamma_{ij}=L^2/\epsilon \, g_{ij}(x, \epsilon)$ on the brane to be normalizable in the asymptotic limit. In what follows we will now show that this contradicts the physical requirement of a massless graviton.

In order to evaluate the relation between normalizable modes and the graviton mass, let us consider a small metric fluctuation 
\begin{align}
    \delta \gamma_{ij}= \gamma_{ij}-\Bar{\gamma}_{ij},
    \label{eq:metricfluctuation}
\end{align}
where $\Bar{\gamma}_{ij}$ is the background metric on the cutoff surface. The dynamics of this graviton excitation is then determined by the linearized Einstein’s field equations on the hypersurface, which in turn is encoded automatically in the mixed boundary condition \eqref{mixedBC}.
At the linearized order, the equations of motion of the graviton excitation read
\begin{align}
-\frac{1}{2}\bigg(1+\frac{(d-1)(d-2)}{2}\bigg) \, \Box \, \delta \gamma_{ij}+ \bigg(\underbrace{\frac{(d-1)(d-2)}{L^2} + \Lambda_{d}}_{\Bar{\Lambda}_{d}}\bigg)\delta \gamma_{ij}=8\pi G_{d} \delta T_{ij}^\mathrm{CFT},
\label{eom}
\end{align}
where we have plugged in the corresponding values of the parameters in general dimensions \mbox{$\kappa_1=\frac{(d-1)L}{16\pi G_{d+1}}$}, $\kappa_2=\frac{d-1}{8\pi G_{d+1}L}$, and used the relation between the Newton constants of different dimensions $G_{d} L = (d-2) G_{d+1}$ \cite{Gubser:1998bc}. 
As such, we can identify a new effective cosmological constant $\Bar{\Lambda}_{d}$ as a combination of the renormalization term governed by $\kappa_2$ and the bare $d$-dimensional cosmological constant, the latter being a part of the local stress-energy tensor on the brane. We are also free to set ${\Lambda}_{d} = 0$ which yields an effective de Sitter cosmological constant.

The term including the perturbation of the stress-energy tensor $\delta T_{ij}^\mathrm{CFT}$ appearing on the right-hand side of \eqref{eom} yields the mass term for the metric fluctuation. If we demand the graviton to be massless, this term has to vanish. We will show that this only happens when the fluctuation contains non-normalizable modes only.

In order to demonstrate this let us further linearize the right-hand side of \eqref{eom} using its expression given in \eqref{TCFTdef}. This yields 
\begin{align}
\begin{split}
    8 \pi  G_\mathrm{N}^{(d+1)} \delta T_{ij}^\mathrm{CFT} &= \frac{1}{2} n^k \partial_k \delta \gamma_{ij}\\ &- \frac{1}{2} n^k (\partial_k \Bar{\gamma}_{nm})\delta \gamma^{nm} \Bar{\gamma}_{ij}
    + \frac{1}{2} n^k (\partial_k \delta \gamma_{nm}) \Bar{\gamma}^{nm} \Bar{\gamma}_{ij} + \frac{1}{2} n^k (\partial_k \Bar{\gamma}_{nm}) \Bar{\gamma}^{nm} \delta \gamma_{ij}.  
\end{split}
\label{deltaTcalc}
\end{align}
Since the graviton on the brane should be a traceless, symmetric rank-2 tensor, the second and third term of (\ref{deltaTcalc}) cancel by virtue of the fact that $\delta \gamma_{nm}\Bar{\gamma}^{nm} = 0$. 
This is a very important point to emphasize. This is the same property that we will use explicitly to derive the propagator for traceless excitations on the brane. The trace part, as we will discuss in the upcoming section, is related to the number of degrees of freedom on the brane. The last term of (\ref{deltaTcalc}) drops out due to the fact that the trace of the induced metric is coordinate independent. Evaluating the extrinsic curvature on the hypersurface $S=\rho - \epsilon = \text{const.}$, the remaining terms read
\begin{align}
   8 \pi  G_{d} \,\delta T_{ij}^\mathrm{CFT}
   = \frac{d-2}{L^2} \left[\partial_\rho (\rho \delta \gamma_{ij}) + \rho (\partial_\rho \delta \gamma_{nm}) \Bar{\gamma}^{nm} \Bar{\gamma}_{ij} - \delta \gamma_{ij}\right],
   \label{deltaTCFT}
\end{align}
The last term only provides a new contribution to the cosmological constant term of Einstein's equations. Bringing this to the left-hand side of \eqref{eom}, this redefines the cosmological constant further as 

\begin{align}
\label{eqn:cosmo-eff}
    \Lambda^{\rm eff}_d = \bar\Lambda_d + \frac{d-2}{L^2} = \frac{(d-1)(d-2)}{L^2} + \Lambda_{d} + \frac{d-2}{L^2}.
\end{align}
Again, using the freedom to choose $\Lambda_{d}$, we can fix $\Lambda^{\rm eff}_d = \frac{(d-1)(d-2)}{L^2}$, the de Sitter value.

The contributions to the graviton mass then come from the first two terms of \eqref{deltaTCFT}. Let us now expand these terms in terms of normalizable and non-normalizable modes decomposed through the Feffermann-Graham expansion of the graviton fluctuation, 
\begin{align}
  \delta \gamma_{ij}(\rho,x) = (\alpha \, \rho^{-1} + \cdots + \beta \, \rho^{d/2-1}) \, h_{ij}(x). 
  \label{modedecomp}
\end{align}
Clearly, $\alpha$ is the non-normalizable mode and $\beta$ is the normalizable mode in this asymptotic expansion. This expansion implies that the tracelessness condition of the graviton holds only if $h_{ij}\Bar{\gamma}^{ij} = 0$. However, since in this expansion $h_{ij}$ only depends on $x$, it is obvious that $(\partial_\rho \delta \gamma_{nm})\Bar{\gamma}^{nm} = 0$ in our case and hence the second term of \eqref{deltaTCFT} vanishes identically.
Inserting the mode decomposition in the single non-vanishing term of \eqref{deltaTCFT} yields
\begin{align}
\label{eqn:RHS-fluctuation}
 \frac{d-2}{L^2} \partial_\rho (\rho \delta \gamma_{ij}) = \frac{d-2}{L^2} \partial_\rho (\alpha  + \beta \, \rho^{d/2}) \, h_{ij}(x) = \frac{d(d-2)}{2 L^2} \beta \, \rho^{d/2} \, \rho^{-1} \, h_{ij}(x) \sim  \frac{d(d-2)}{2 L^2} \frac{\beta}{\alpha} \rho^{d/2} \delta \gamma_{ij}.
\end{align}
In the last step, we used the fact that the non-normalizable part of the graviton fluctuation dominates near the position of brane at $\rho=\epsilon$ close to the boundary, namely, $\delta \gamma_{ij} \sim \alpha \rho^{-1} h_{ij}$. 
In the near boundary expansion $\epsilon \rightarrow 0$ one can read off the graviton mass directly from \eqref{eqn:RHS-fluctuation} as
\begin{align}
m_{\rm grav}^2 = - \frac{\beta}{\alpha} \frac{2\, d (d-2)}{(d-1)(d-2)+2} \frac{\epsilon^{d/2}}{L^2}.
\label{massterm}
\end{align}
As is evident from \eqref{massterm}, in order to get a massless graviton, one does need to set $\beta = 0$, i.e. we need to exclude the normalizable mode from the field expansion. {\it In other words, in order to get a consistent holographic description of such a cosmological model, one needs to consider only non-normalizable boundary conditions.} This conclusion is very much in line with that of \cite{Neuenfeld:2021wbl}.

As mentioned in the previous section, there should also be additional trace contributions to the CFT stress-energy tensor.  By construction, such contributions are normalizable, as can be verified directly by performing a near boundary expansion $a \rightarrow \infty$ of \eqref{eqn:epsilon-p}. 
However, this anomalous part of the stress-energy tensor does not contribute to a mass term, rather it provides the number of degrees of freedom imprinted on the cutoff brane. We will come back to this, once again, in the context of our dark bubble model in the upcoming section.

\section{From braneworld to bubbleworld: The graviton propagator on the shell} \label{sec:bubble}
So far we have discussed a simple one-sided case with an $\text{AdS}_{d+1}$ bulk geometry with a cutoff brane near the boundary. Now we extend this configuration, replacing the cutoff with a spherical shell having an interior as well as an exterior region. This is the dark bubble model set up and developed in \cite{Banerjee:2019fzz, Banerjee:2020wix, Banerjee:2020wov, Banerjee:2021qei, Banerjee:2021yrb, Banerjee:2022myh,Danielsson:2021tyb, Danielsson:2022lsl,Danielsson:2022fhd,Basile:2023tvh}. In this model, the AdS spaces in the inside $(-)$ and outside $(+)$ have different length scales with $\Lambda_+ > \Lambda_-$ which ensures the expansion of the dark bubble. The junction across the shell, separating the two regions, determines the tension of the bubble wall. As discussed before, a subcritical tension of the wall leads to a positive cosmological constant on it, corresponding to the desired de Sitter universe. The non-identical AdS spaces in the interior and exterior region render the gravitational constant finite, but for this one does need to allow non-normalizable modes in these regions, which in turn imprint massless graviton modes on the bubble wall. In the dark bubble model such non-normalizable modes are physically realized using hanging strings as sources. These provide the required kinetic energy for the bubble to expand in the presence of particles of dust represented by the end points of the strings \cite{Banerjee:2020wov}.

Our goal in this subsection is to fit the dark bubble model into the holographic formalism discussed in the previous sections. The major technical difference is how to incorporate the junction condition across the bubble wall that separates the two holographic bulk spacetimes. The location of the bubble wall coincides with the near boundary regions of both the bulk regions. In this setup the graviton propagator on the bubble was computed in \cite{Banerjee:2020wix}. We will revisit this computation in this section, but, this time, in the light of the holographic formalism developed in the previous sections. 

Unlike the previous section, where the brane is merely a cutoff near the boundary, the dynamics of the bubble wall plays a more crucial role in the dark bubble model.  
As discussed above, in order to implement the junction between interior and exterior with a view to get an expanding universe on the bubble, we require a subcritical positive tension spherical brane to be realized as the bubble wall. The latter modifies the bulk action as 
\begin{align}
S = S_\mathrm{\rm{bulk}} + S_\mathrm{\rm{brane}} = S_\mathrm{\rm{bulk}} + \sigma \int \mathrm{d}^d x \sqrt{-\gamma} \, ,
\label{bubblefullaction}
\end{align}
where $\sigma$ is the brane tension.

 The extra contribution due to the brane only depends on the determinant of the induced metric and is of the same form as that of the constant term proportional to $\kappa_2$ in the counterterm action \eqref{countertermaction} in the bulk.
 Therefore, in the equation of motion, this will provide a new contribution to the cosmological constant term. The derivation is similar to that leading to \eqref{eqn:cosmo-eff}, however, there are a couple of interesting catch points, particularly for the dark bubble model, which has an interior and an exterior. The latter is manifest in the difference of the AdS scales appearing in the denominator of the RHS of \eqref{eqn:G4-DB}. This has an interesting consequence in terms of a different hierarchy between the four-dimensional and the five-dimensional Newton's constants given by \cite{Danielsson:2022lsl,Danielsson:2023alz}
 \begin{equation}
 \label{eqn:G4-G5-new}
 \frac{G_4}{G_5} = -\frac{3}{L} \frac{N}{\Delta N},
 \end{equation}
 where $\Delta N$ is a negative integer signifying 
 the number of nucleated D$3$ branes from a stack of N D$3$ branes in the ten-dimensional realization of the dark bubble model in type IIB supergravity \cite{Danielsson:2022lsl}. This new scaling behaviour follows precisely due to the difference in scales appearing in the denominator of \eqref{eqn:G4-DB}, rather than the sum, as usually for RS-like braneworld models. The second important point is that we need to run the holographic renormalization scheme from either side of the brane consistent with the junction condition. This would fix the renormalization parameters as \begin{equation}\label{eqn:eqchoice} \kappa_1 = \frac{3 \left(L_+ - L_-\right)}{ 16\pi G_5 } \ \ , \ \ \ \ \kappa_2 = \frac{3}{8\pi G_5} \left(\frac{1}{L_-} - \frac{1}{L_+}\right), \end{equation}
 for the two-sided geometry. 
Taking these two special features of the dark bubble model into account, along with the contribution of the brane tension, we end up in getting the effective four-dimensional cosmological constant as\footnote{We fix the contribution to the cosmological constant coming from the local term exactly as before, namely, by choosing $\Lambda_4$ in such a way that it cancels the contribution coming from the CFT stress-energy tensor.}
 \begin{align}
\label{eqn:cosmo-eff-branes}
    \Lambda^{\rm eff}_4 = 8 \pi G_4 \kappa_2 - \sigma = \frac{6}{L^2} - \sigma.
\end{align}
It is interesting to note that, at the end of the day, the effective cosmological constant retains its form as in the standard one-sided holography, as in \eqref{eqn:cosmo-eff}. Nevertheless, the scale of hierarchy, $N$ appearing in \eqref{eqn:G4-G5-new}, makes our dark bubble significantly different from the inside-inside braneworld models, including the RS. This difference provides a reversal of hierarchy beteen the five-dimensional and four-dimensional Planck scales, resulting in interesting phenomenological implications. Interested readers are referred to \cite{Danielsson:2023alz} for a detailed account of the latter.

Note, that this is precisely the desired cosmological constant \eqref{ccbubble} on the brane, with the critical tension of the brane holographically identified with the parameter of holographic renormalization, $8 \pi G_4 \kappa_2$. Varying the full action \eqref{bubblefullaction} yields the equation of motion  
\begin{align}
\mathcal{G}_{\mu \nu}=\frac{1}{2 M^{d-1}}[T_{\mu \nu}-\Lambda g_{\mu \nu} - \sigma (g_{\mu \nu} - n_{\mu} n_{\nu})\delta(X^{d+1}-X^{d+1}(x))],
\label{Einsteintensor}
\end{align}
where $\mathcal{G}_{\mu \nu}$ is the Einstein's tensor, and $n_{\mu}$ is the normal vector on the brane. The delta function appearing in the field equation selects the bulk position of the shell parametrized by its world-volume coordinates $x^a$.

Our strategy is first to linearize the equation of motion in terms of the leading graviton fluctuation as before. For this, it turns out to be convenient to use the Gauss normal coordinates, in which the AdS metric assumes the form \cite{Banerjee:2020wix} 
\begin{align}
ds^2 = d\xi^2 + a^2(\xi) (\eta_{ab} + h_{ab}\left(\xi, x^a)\right) dx^a dx^b.
\end{align}
One can obtain this metric from the Fefferman-Graham metric \eqref{modedecomp} simply by rescaling the radial coordinate as 
\begin{align}
\rho \propto e^{-2k\xi} = a^{-2}(\xi),
\label{eqn:rho-a-trafo}
\end{align}
with $k = \frac{1}{L}$ being the inverse of the AdS length scale.
 
The main algebraic difference we have here, as compared to the simpler one-sided braneworld construction presented in the previous section, is the last term in \eqref{Einsteintensor}. It contains products of normal vectors on the brane which give additional derivative terms with respect to the normal coordinate when we linearize the equations of motion in the metric fluctuation $h_{ab}$.
At the leading order of the metric perturbation in $h_{ab}$, the transverse components of \eqref{Einsteintensor} yields \cite{Giddings:2000mu}
\begin{align}
\begin{split}
\Box \Bar{h}_{ab} = e^{2k\xi}(&-\eta_{ab} \partial^n \partial^m \Bar{h}_{n m} + \partial^n \partial_a \Bar{h}_{b n} +  \partial^n \partial_b \Bar{h}_{a n})\\
&+ \frac{\eta_{ab}}{2} e^{k \xi d} \partial_\xi (e^{-k\xi d} \partial_\xi {\bar h}) - 16 \pi G_{d+1} e^{2k\xi} T_{ab},
\end{split}
\label{EFElinearized}
\end{align}
where we have defined the trace-reversed fluctuations
\begin{align}
\Bar{h}_{ab} = h_{a b} - \frac{1}{2}\eta_{a b} h
\end{align}
satisfying the transverse gauge condition $\partial^a \Bar{h}_{ab} =0$ everywhere outside the location of the source \cite{Giddings:2000mu}.
Absorbing the scale factor by defining $\gamma_{ab}:=a^2(\xi) h_{ab}$ and taking a trace, one obtains from \eqref{EFElinearized} a trace equation in $d=4$ of the form
\begin{align}
a^{-2}\partial^2\Bar{\gamma} + 3 (\partial^2_\xi - 4k^2)\Bar{\gamma}=-16\pi G_5 T,
\label{eqn:trace}
\end{align}
where $\bar\gamma_{ab}$ is the trace reversed $\gamma_{ab}$.

We need to subtract \eqref{eqn:trace} from the original linearized equation \eqref{EFElinearized}, in order to select the traceless graviton mode. With this, finally, the equations of motion for the graviton on the dark bubble assume the form \cite{Banerjee:2020wix}
\begin{align}
\Box \chi_{ab} + (\partial_\xi^2-4k^2)\chi_{ab} = -16 \pi G_5 \Sigma_{ab},
\label{eombubble}
\end{align}
where $\chi_{ab}$ and $\Sigma_{ab}$ represent the traceless metric excitation and stress tensor, respectively.

Before we solve the equations of motion and continue with the derivation of the propagator, let us pause a bit to check the role of normalizable and non-normalizable modes in generating the mass of these transverse traceless modes. We follow the same strategy as discussed in section \ref{sec:NN}, namely to study the equations of  motion of these modes near the boundary to extract the mass term. In particular, it would be interesting to see whether the additional piece of the dark bubble equations of motion \eqref{eombubble}, arising due to the shell tension, alters the previous conclusion in any way.  
Following this motivation, we will concentrate on the second term in \eqref{eombubble}, since the stress tensor contribution is the same in both cases.

To accomplish our goal, we need to first go back to the Fefferman-Graham radial coordinates through \eqref{eqn:rho-a-trafo}. In this coordinate, 
the relevant piece of the equations of motion transforms into the form
\begin{align}
(\partial_\xi^2-4k^2)\chi_{ab} = 4k^2[-\partial_\rho (\rho \chi_{ab}) + \rho \partial_\rho^2 (\rho \chi_{ab})].
\end{align}
Once again, we plug in this the mode decomposition of the metric fluctuation \eqref{modedecomp} for $d=4$. It yields
\begin{align}
(\partial_\xi^2-4k^2)\chi_{ab} 
= 4k^2(-2\beta \rho+2\beta \rho)h_{ab}=0. 
\end{align}
Thus, taking the limit $\rho \rightarrow a_s^{-2}$, with $a_s \rightarrow \infty$ being the location of the brane, we obtain a vanishing correction to the mass term at the leading order. Therefore, in the $d=4$ holographic dark bubble, we end up getting, exactly, the same mass term of the graviton \eqref{massterm} on the bubble wall \footnote{The mass terms get modified in higher dimensions. Nevertheless, they remain proportional to the ratio between the normalizable and the non-normalizable modes. Consequently, the conclusion remains unaltered.}. Accordingly, the condition for having a massless graviton on the brane remains the same as well, i.e., in the holographic model of dark bubble, we do need to turn off the normalizable fluctuation modes and should only allow non-normalizable modes instead. As discussed in the introduction, physically, this condition fits perfectly with our dark bubble model, as the non-normalizable modes in this model are carried by hanging strings present in this model. A detailed discussion on this can be found in \cite{Banerjee:2020wov}.

We now carry forward this finding towards determining the real dynamics of the graviton on the bubble wall. For deriving the propagator on the wall, it is easier to work in momentum space. We therefore apply a transverse Fourier transformation to \eqref{eombubble} that yields \cite{Banerjee:2020wix}
\begin{align}
\label{eqn:eom-chi}
 \bigg(- \frac{p^2}{a^2} + \partial_\xi^2 - 4 k^2\bigg) \Tilde{\chi}_{ab}(p,\xi)=-16\pi G_5 \Tilde{\Sigma}_{ab}.
\end{align}
 $\Tilde{\chi}_{ab}$ and $\Tilde{\Sigma}_{ab}$ are the transverse Fourier transforms of the trace removed metric and stress tensor, respectively. The shell is assumed to be localized at $a_s\equiv a_+=a_-$, where $a_\pm$ describes the scale factor in the interior or exterior region.  It is instructive to first solve for a massless minimally coupled scalar propagator in AdS. The latter essentially satisfies the same equation of motion 
 as \eqref{eqn:eom-chi}, and reproduces all the relevant qualitative features of the solutions of \eqref{eqn:eom-chi}. Finally, of course, we will reinstate the indices properly to write down the solution for the graviton mode explicitly. The full computation was done in \cite{Banerjee:2020wix}. We will here only show the main steps and make it aligned with the holographic lessons we learnt above.

We start with an ansatz for a solution of the Green's function for the exterior and interior regions 
\begin{align}
 \Delta^+_{\Tilde{\chi}} (p;a_+,a_-)&=A(p,a_-)K_2\bigg(\frac{p}{k_+a_+}\bigg)+B(p,a_-)I_2\bigg(\frac{p}{k_+ a_+}\bigg)\\
 \Delta^-_{\Tilde{\chi}} (p;a_+,a_-)&=C(p,a_+)K_2\bigg(\frac{p}{k_-a_-}\bigg),
 \label{innerpropagator}
\end{align}
valid outside the source, which in this case is the brane at the junction.
$K_2$ and $I_2$ denote the respective modified Bessel functions, and represent non-normalizable and normalizable metric modes in the large $a$ limit. The normalizable piece proportional to $I_2$ does not appear in \eqref{innerpropagator} because it is divergent in the limit $a\rightarrow 0$ where we expect our solution to be regular.
We fix two of the coefficients $A, B, C$ using junction conditions on the bubble wall. There are two matching conditions, one for the propagator function and the other involving its derivative, namely
\begin{align}
\Delta^-_{\Tilde{\chi}}(p;a_+,a_s)=\Delta^+_{\Tilde{\chi}}(p; a_s,a_-),\\
\frac{1}{16\pi G_5} \bigg \lbrack \frac{\partial}{\partial \xi_i}\Delta^i_{\Tilde{\chi}}(p;a_+,a_-)|_{a_i \rightarrow a_s} \bigg \rbrack^{i=+}_{i=-} + \frac{\sigma}{3}\Delta^+_{\Tilde{\chi}}(p; a_s,a_-) = \frac{1}{16\pi G_5}.
\end{align}
Solving this, we express $A$ and $C$ in terms of $B$, which leads to a Green's function of the form
\begin{align}
\label{eqn:propagatorDelta}
\Delta^+_{\Tilde{\chi}} (p;a_+,a_-) = &-\bigg\lbrack\frac{1}{g_K(p,a_s)}- B(p,a_-) \frac{g_I(p,a_s)}{g_K(p,a_s)} \bigg\rbrack K_2\bigg(\frac{p}{k_+ a_+}\bigg)\\
&+ B(p,a_-) I_2\bigg(\frac{p}{k_+ a_+}\bigg),
\end{align}
with
\begin{align}
g_K(p,a_s) &= \frac{p}{a_s} \frac{K_2\bigg(\frac{p}{k_+ a_s}\bigg) K_1\bigg(\frac{p}{k_- a_s}\bigg) - K_2\bigg(\frac{p}{k_- a_s}\bigg) K_1\bigg(\frac{p}{k_+ a_s}\bigg)}{K_2\bigg(\frac{p}{k_- a_s}\bigg)},\\
g_I(p,a_s) &= \frac{p}{a_s} \frac{I_2\bigg(\frac{p}{k_+ a_s}\bigg) K_1\bigg(\frac{p}{k_- a_s}\bigg) + K_2\bigg(\frac{p}{k_- a_s}\bigg) I_1\bigg(\frac{p}{k_+ a_s}\bigg)}{K_2\bigg(\frac{p}{k_- a_s}\bigg)}.
\end{align}
{\it This is the time to use our holographic understanding of boundary conditions as we now aim to fix the last remaining function, $B(p,a_s)$ on the brane.} We recall the requirement for having a massless graviton on the wall, in the limit when the brane is close to the asymptotic boundaries of either AdS spacetime. In the context of the present model, this demands setting the normalizable mode to be set to zero in the near boundary expansion $a_s \rightarrow 0$ of \eqref{eqn:propagatorDelta}. This in turn fixes $B(p,a_s)$ as 
\begin{align}
\label{eqn:B-fix}
B(p,a_s) = =-\frac{\eta}{\eta g_I(p,a_s) - 4 g_K(p,a_s)} \ \ \text{with} \ \ \eta:= 3 - 4\gamma + 4 \ln{2}.
\end{align}
Plugging \eqref{eqn:B-fix} in \eqref{eqn:propagatorDelta} in the low momentum limit yields the propagator on the bubble wall as 
\begin{align}
\Delta^s_{\Tilde{\chi}}(p;a_s,a_s)=\frac{a_s^2}{p^2}\bigg(\frac{2k_-k_+}{k_- - k_+}\bigg) + \mathcal{O}(p^0).
\end{align}
Finally, after convoluting the Green's function with the source term which corresponds to a simple multiplication in momentum space, the Fourier component of the graviton mode on the bubble wall is
\begin{align}
\label{eqn:chi-propagator}
\Tilde{\chi}^s_{ab}(p;a_s)=-16\pi G_5 \frac{a_s^2}{p^2}\bigg(\frac{2k_-k_+}{k_- - k_+}\bigg) \Tilde{\Sigma}_{ab}.
\end{align}

This recovers, in the small momentum limit, the expected four-dimensional graviton mode as imprinted on the bubble wall with the correct four-dimensional Newton's constant \cite{Banerjee:2018qey}. 
{\it The key point we want to emphasize in this paper, through the computation leading to \eqref{eqn:chi-propagator}, is the role of holography in reproducing the correct propagator on the bubble wall.} We use the requirement of having a massless graviton on the brane, which automatically dictates the proper boundary condition. Surprisingly enough, this was already an energetic requirement in our dark bubble model to ensure an expanding universe on the wall \cite{Banerjee:2018qey, Banerjee:2019fzz}. The present work reestablishes it in terms of a proper holographic requirement.

We will conclude this section with a pending discussion on the degrees of freedom on the brane. As outlined in section \ref{sec:NN}, the CFT stress-energy tensor generally has a trace piece due to the conformal Weyl anomaly that we subtracted from the equations of motion in order to take into account the tracelessness of the graviton imprinted on the bubble wall. We now want to emphasize the physical connection of the trace piece to the number of degrees of freedom on the wall. Similarly to section \ref{sec:NN}, we need to consider the stress-energy tensor of the CFT which is renormalized by means of the holographic renormalization techniques \cite{Skenderis:2002wp}. In the two-sided case of the dark bubble, the choice of the renormalization parameters \eqref{eqn:eqchoice}
renders finiteness of the brane stress-energy tensor in the limit $a_s\to\infty$.
  
We now compute the trace of the regularized stress energy tensor, and compare it to the expected Weyl anomaly for the conformal field theory on the brane, namely, \cite{Henningson:1998gx, Balasubramanian:1999re}
 
 \be
 \text{Tr} \, T = -\frac{N_{\text{eff}}^2}{32 \pi^2} \left(-{\cal R}_{ij}{\cal R}^{ij} + \frac{1}{3}{\cal R}^2\right),
 \ee 
which yields, for our case, 
  \be
  N_{\text{eff}}^2 = \frac{\pi}{2 G_5} \left(L_+^3 - L_-^3\right)  .
  \ee
This matches perfectly with the thermodynamical derivation presented in \cite{Banerjee:2018qey}. Usually, when we simply have one term of type $\frac{\pi L^3}{2 G_5}$, this translates into $N^2$, which is the number of adjoint fields. For the dark bubble, we find, due to the subtraction, $N$, which is the dimension of the fundamental representation. This was also used in \cite{Danielsson:2022lsl}, when discussing corrections to the string tension. The physical picture is that of the background $AdS_5 \times S^5$, dual to $N$ branes. In this background a single brane nucleates, the dark bubble. The counting is that of open strings connecting the brane to the background. It is interesting to note that the number of degrees of freedom is always $\sim L^2/G_4$. 

\section{A brief summary of the main claim}
\label{sec:conclusion}
The aim of this work is to fit the dark bubble model consistently within a holographic framework \cite{Banerjee:2019fzz, Banerjee:2020wix, Banerjee:2020wov, Banerjee:2021qei, Banerjee:2021yrb, Banerjee:2022myh,Danielsson:2021tyb, Danielsson:2022lsl,Danielsson:2022fhd,Basile:2023tvh}. We manage to achieve this through a careful and thorough investigation of a broader class of boundary conditions allowed in AdS/CFT \cite{Compere:2008us}. While such constructions were already done in the context of holographic cosmology \cite{Apostolopoulos:2008ru, Banerjee:2012dw}, the challenge we faced here was to incorporate the two-sided bulk geometry, living on either side of the bubble wall, which, geometrically, are the inside and the outside of the bubble. A second challenge was to make the dark bubble model phenomenologically reasonable to produce the correct dynamics of the graviton on the brane, along with a finite four-dimensional Newton's constant. We noticed the necessity of having extended objects like strings carrying momentum, which in turn imprints matter on the bubble wall. Not only that, these strings also ensure proper temporal evolution of the bubble, which from the perspective of the physics on the wall can be realized through a process of mass renormalization (see \cite{Banerjee:2020wov} for a detailed discussion on it). The latter ensures finite masses of the particles induced on the wall. 

While we needed the strings to ensure the consistency of the model, it was a challenge to fit this aspect in the context of holography. In this paper, we showed that this requirement can be converted into the requirement of having a massless induced graviton on the bubble wall. For this we need to discard the normalizable components of the asymptotic expansion of bulk graviton fluctuation keeping only the non-normalizable piece. A similar conclusion was drawn recently in \cite{Neuenfeld:2021wbl} in the context of a one-sided Karch-Randall braneworld scenario. In our two-sided case, this requirement fits nicely, in the sense that these non-normalizable modes are nothing but an artifact of the momentum carrying strings that makes the dark bubble a suitable model of the expanding universe, while being perfectly consistent with the swampland conjectures avoiding the usual no-go theorems that prevent generating a de Sitter universe imprinted on a conventional positive tension braneworld \cite{Banerjee:2021yrb}. This is also where our construction significantly differs from the Randall-Sundrum braneworld story \cite{Banerjee:2022myh}. We revisited the computation of the graviton propagator on the brane \cite{Banerjee:2020wix}, but now endowed with the complete holographic understanding of the conditions to be imposed to fix the relevant functions on the brane. It yields the correct massless graviton propagator in the low energy limit, with the correct four-dimensional Newton's constant.

Last but not least, in this process of rediscovering our model in the light of holography, we interpret the critical tension of the brane in terms of a parameter of holographic renormalization and also compute the effective degrees of freedom on the bubble wall from the conformal Weyl anomaly of the holographically renormalized stress-energy tensor. The degrees of freedom matches exactly with its thermodynamical estimation presented in \cite{Banerjee:2018qey}.

 \acknowledgments
 We would like to thank Johanna Erdmenger, Suvendu Giri, Andreas Karch, Dominik Neuenfeld, Daniel Panizo and Thomas Van Riet for useful discussions and communications. We would also like to thank Dominik Neuenfeld for comments on an earlier version of the manuscript.

\bibliography{references}
\bibliographystyle{JHEP}

\end{document}